\documentstyle[preprint,aps]{revtex}
\preprint{Applied Physics Report 94-40}
\input psfig
\begin{document}
%\preprint{Version by Peter, 11/08/94 revised by Peter 26/08/94 \\
%Revised by Yuri, 23/08/94, Back from Peter 29/08/94, Returned by Yuri
%%29/08/94}
%\draft
\title{\bf
%\vspace*{-10mm}
%\begin{flushright}
%{\large Applied Physics Report 94-40}
%\end{flushright}
%\vspace*{5mm}
%%%%%%%%%%%% PH Maybe 'gate voltage induced' give wrong associations
%%%%%%%%%%%% to the {\em applied} gate voltage, what's your opinion?:
%%%%%Gate Voltage Induced
%%%Low Frequency
% Yuri's suggestion:
Flicker Noise Induced by Dynamic Impurities in a Quantum Point Contact}
\author{J. P. Hessling\thanks{Electronic address:
hessling@fy.chalmers.se}}\address{
Department of Applied Physics, Chalmers University of
Technology and G\"{o}teborg University, \\
S-412 96, G{\"o}teborg, Sweden}
\author{Yu. M. Galperin}
\address{
Department of Physics, University of Oslo,
P.O. Box 1048 Blindern, N 0316 Oslo 3, Norway, \\
and A.F. Ioffe Physico-Technical Institute, 194021 St. Petersburg,
Russia\\
\mbox{\rm \today}}
%\\
%\vspace*{5mm}

\maketitle

\begin{abstract}
%\parbox{14cm}{\medskip \rm \indent
We calculate low-frequency noise (LFN) in a quantum point contact (QPC)
which is
electrostatically defined  in  a  2D  electron  gas of a GaAs-AlGaAs
heterostructure. The conventional source of LFN in such
systems are scattering potentials fluctuating in time acting
upon injected electrons. One can discriminate between potentials
of different origin -- noise may be caused by the externally applied
gate- and source-drain  voltages, the motion of defects
with internal degrees of
freedom close to the channel, electrons hopping between localized
states in the doped region, etc.
In the present study we propose a model of LFN based upon the
assumption that there are many dynamic defects in the surrounding
of a QPC.
A general expression for the time-dependent current-current
correlation function is derived and applied to a QPC with quantized
conductance. It is shown that the level of LFN is significantly different
at and between the steps in a plot of the conductance vs. gate voltage.
On the plateaus, the level of noise is found to be low and strongly
model-dependent.
At the steps, LFN is much larger and only weakly model-dependent.
As long as the
system is biased to be at a fixed position relative the
conductance step,
we find that the level of noise is independent of the number
of conducting modes.
{}From numerical calculations we conclude that the level of noise approximately
obeys a power law as a function of frequency
for frequencies larger than a threshold.
At the steps for frequencies larger than
the minimal transition
rate for the dynamical impurities, we have
$S(\omega) \propto 1/\omega^{0.85}$.
We are convinced that noise measurements will play a crucial role
in the course of investigating the effect of the environment in QPCs.
%\pacs
\end{abstract}

\narrowtext
\section{Introduction}

%%%%%%%%%%%%%%%%% FIG. ALGAAS.EPS %%%%%%%%%%%%%%%%

Electron transport through % PH: Why? so-called
ballistic quantum
point contacts  (QPC), shown in Fig. ~\ref{fig:gaas},
has been extensively studied during the last decade
(for a review see Ref.~\onlinecite{BeenHout}).
One of the most interesting properties of these systems is the
quantization  \cite{wees,kwh,wharam}  of the
conductance as a function of the
applied gate voltage  (for a review see Ref.
\onlinecite{BeenHout} and references therein).
%% PH: quantization -> nonlinear character why say it again?
%%     probably you mean small oscillation on 'top'?
In addition, small oscillations
in the $I-V_{sd}$   curves have been found.
%as well as
%nonlinear   character   of   their  $I-V$   curves.
The quantum channel is assumed to behave as a wave guide where
the number of transverse modes
is  dependent  on the  gate  voltage $V_g$ as well as on the source-drain
voltage  $V_{sd}$. According to the present understanding (see e.g.
Ref.~\onlinecite{BeenHout}), the steps in the conductance when measured
vs. gate voltage are  due to switching of the effective number of transverse
modes in the channel.
The small-scale oscillations in the $I-V_{sd}$-curves are also
believed to be caused by this kind of
switching
\cite{gh,zag1,pepper}.
%%PH Already said above: of transversal modes.
%The above mentioned switching of the transversal modes
%number is also responsible for nonlinear character of $I-V$ curves.

Low frequency noise in QPCs may originate from time-dependent
internal as well as
external parameters. Fluctuations in the potentials applied to the electrodes
are examples of external sources of noise. Internal noise
can for instance be generated by the motion of impurities within the contact
region, or by a time-dependent rearrangement of charge in the doped
region of the structure.

Here we focus on noise of internal origin which is
intrinsic for any realistic structure.
It is well established that there is some disorder in the vicinity of
small devices   even if they are of high quality.
In any disordered system, defects with internal degrees of freedom are present
which may change their position due to
interactions with a thermal bath.
There will hence be a time-dependent
impurity potential.
These defects usually switch between two states, leading to a
telegraph-like noise.
Dynamical defects of this type were observed by several authors in
the surrounding of even very high quality classical point contacts
\cite{buhrman1,buhr,buhrman3,buhrman4,zimmerman,golding}.
The role of the  defects  with  internal  degrees  of  freedom  in
the surrounding of point
contacts has been  extensively discussed in the literature (for a review see
Ref. \onlinecite{gkk}   and  references   therein).
Recently,  new  features have been studied \cite{hdw,mwl1,mwl2}
which are due to the Kondo effect
in scattering by two-level impurities.
%Also electron-electron interactions.

It is quite reasonable to believe that such %fluctuating
defects,
so-called {\it elementary fluctuators} (EFs), also exist in
the surrounding of ballistic quantum point contacts.
There are experimental
results for QPCs which can be
explained by the presence of defects with internal
degrees of freedom. In particular,
we have made an attempt
\cite{hessgalp1} to find theoretical support for
experimentally observed correlation between
telegraph-like noise and current  in QPCs
\cite{lin3}.
An oscillating response to small  source-drain voltages
was found
\cite{lin,lin2}.

The  microscopic structure of the EFs is not yet completely clear.
One of the proposed
explanations for the presence of
two-state  defects  is  disorder-induced  soft  atomic
vibrations.   For  low  excitation  energies  the  vibrations  are  strongly
anharmonic and can be described as an atom or group of atoms moving  in  an
effective  double-well  potential.  Such  entities  are  known  as  two-level
tunneling systems (TLS) \cite{and,phil}. These are responsible for the low
temperature properties of glassy materials. The generalization of the
TLS  model  for  higher  excitation  energies  was worked out in Ref.
\onlinecite{karp}.   Dynamical   defects   produce  %elastic
electric
fields,  slowly   varying  in time. The fields will
scatter conduction electrons, thereby
creating a sort of `noisy environment'.
Another possibility for having fluctuations in the scattering
potential is electrons hopping
between adjacent impurity centers in the doped region.

Our aim is to analyze low-frequency noise (LFN) in the current, caused by
a `noisy environment'  of a QPC. In this study we restrict ourselves
to systems with a wide spacer between the doped region
and the 2DEG where the quantum channel is defined, see Fig. ~\ref{fig:gaas}.
If the spacer is wide enough (which is the case for high-mobility
structures), the correlation length of the impurity potential
% produced
%by the doped region with a relatively large
%degree of disorder
may exceed the size of the QPC. Hence, the
impurities in the doped region will contribute to the smooth static
potential felt by the electrons in the quantum channel. A time-dependent
rearrangement of the charge in the doped region leads to
almost homogeneous fluctuations of the potential in space
even if the voltages between the gate electrodes are kept
constant. Effectively, these fluctuations manifest
themselves as variations in time of the effective gate voltage.
% which
%enters the theory of ballistic transport (see e. g.
%Ref.~\onlinecite{BeenHout}).

At the plateaus in conductance %quantization with gate voltage
such fluctuations cannot change the current considerably.
However, close to the steps transverse modes may be turned
on or off.
%%PH: Unnecessary repetition: by these fluctuations.
As a result, the level of LFN should
increase significantly in the  vicinities of the conductance steps.

The conductance of ballistic channels is a strongly non-linear property.
Hence one can expect to find rather unusual and informative
properties of LFN of such contacts.
Experimental studies %%%%%%of the statistical
%%%%%%properties
may serve as tools for extracting
the principal interaction mechanisms between
electrons in QPCs and their environment.
Studies of this kind may  also allow us to draw conclusions about
the spatial distribution as well as the intrinsic dynamics of
the impurities.

This article
 % WHY is this not a good word???? personally I STRONGLY believe that
 % paper is slang...
 % that paper is commonly used is not a valid argument to me!
is  organized  as follows.
In Section \ref{curr_section}, the current through an adiabatic contact
is discussed which provides us with the starting point
of our derivation of the level of noise.
The model for  low-frequency noise (LFN) in a QPC is formulated in Section
\ref{noise}. In this model, LFN is caused by EFs distributed
randomly in the vicinity of a QPC.

General analytical  expressions for low-frequency
noise
%, which we describe as  caused by a set of randomly distributed in
%space EFs,
as well as results from numerical calculations,
will  be  given  in Section \ref{noise}.
In Section \ref{discussion} we discuss our main results and draw conclusions.
We have also submitted a
derivation in Appendix \ref{app:samenoise}
explaining why, under some circumstances,
noise is independent of the number of propagating modes.

\section{Model} \label{curr_section}

The system under study consists of an
adiabatically   smooth    channel  \cite{glks}
connecting  two  equilibrium reservoirs. It is formed in a 2D electron
gas  by the split gate technique. The gates are assumed to provide
a hard wall potential in the transverse direction ($y$),
slowly (adiabatically) varying along the channel ($x$).
Accordingly, the WKB approximation for the electron wave function is
applicable. This wave function can be written as \cite{glks}
\begin{equation} \label{WKB-WF}
\Psi(x,y,E)  = \sum_{n,\pm} a_{n}^{\pm}\chi_{n}^{\pm}(x,y,E),
\end{equation}
where
\begin{eqnarray}
\chi_{n}^{\pm}(x,y,E)
& = &\sqrt{k_{n,\parallel}(E,\mp \infty) / k_{n,\parallel}(E,x)} \phi_{n,x}(y)
\\
& \times & \exp\left[i\int_{\mp \infty}^x dx^\prime
k_{n,\parallel}(E,x^\prime)\right], \label{not} \\
k_{n,\parallel}(E,x)
& = &k_{\text F} \sqrt{\varepsilon-\varepsilon_{n,\perp}-\Upsilon(x,t)}.
\label{kl}
\end{eqnarray}
Here $E\equiv \varepsilon E_{\text F}$ is the total energy of the electron ---
$E_{\text F}$ being  the Fermi
energy in the leads at zero bias voltage, $V_{sd}=0$ --- while
$k_{\text F}=\left({2mE_{\text F}/\hbar^2}\right)^{1/2}$
is the Fermi wave vector and
$k_{n,\parallel}(E,x)$ the longitudinal wave vector along the channel.
The transverse
part of the wave function, $\phi_{n,x}(y)$, depends parametrically on the
longitudinal coordinate $x$; the corresponding `transverse' energy eigenvalue
is
$\varepsilon_{n,\perp}$ and is measured in units of the Fermi energy $E_{\text
F}$.
Hence,
\begin{equation}
\varepsilon_{n,\perp}= \left(\frac{\pi n }{k_{\text F} d}  \right)^{2},
\end{equation}
where  $d$  is the width of the  channel, for simplicity assumed to be
constant.

The plus sign ($+$) corresponds to propagation from the left to the  right
reservoir while the minus sign ($-$)
corresponds to transmission in the opposite direction.

The dimensionless function $\Upsilon (x,t)$ is the total
potential in units of $E_F$
felt by electrons moving in the electric field produced by the electrodes
and the
impurities in the doped region.
We specify $\Upsilon$ as a sum
$
\Upsilon (x,t) = u_0 (x) + \delta u (x,t) + v s(x)
$
where $u_0 (x) \approx eV_g/E_{\text F}$ comes from the static
part of the gate voltage.
The fluctuating contribution from
impurities and defects in the doped region is modeled by
$\delta u (x,t)$, while $v$ is the
potential caused by the source-drain voltage
($v = eV_{sd}/2E_{\text F}$).
In order to match the Fermi levels in both reservoirs to the left and to the
right, we  must require that $s (\pm \infty) = \pm 1$.
To simplify we have considered the case of small source-drain
voltages and put $s(x)=0$.
We also neglect the $x$-dependencies of the potentials
$u_0$ and $\delta u$ giving
\begin{equation} \label{kl2}
\Upsilon (x,t) = u_0 + \delta u (t) .
\end{equation}
For convenience we introduce a dimensionless parameter $\eta$,
\begin{equation}\label{n0}
[n/\eta (u_0)]^2 \equiv \varepsilon_{n,\perp}-u_0.
\end{equation}
The use of the parameter $\eta$ allows for a
particularly simple interpretation
in terms of the well known conductance quantization with gate voltage
 \cite{glks}.
If $\eta$ is an integer, say $m$, mode number $m$ is just about to
start conducting.  Small fluctuations
in the gate and source-drain voltages may then induce
variations in the conductance  between $m-1$ and $m$ in units of the
fundamental conductance quantum $2e^2/ h$. When $\eta= m +0.5$
we are exactly half way between two adjacent steps in
the conductance when it
is plotted versus gate voltage. From now on, $u_0$ is removed.
The offset gate voltage $u_0$ which before affected the
bottom of the potential well in which the electrons move, is now
instead included
in the effective width $d$ of the channel.

Note that according to our model the fluctuating part $\delta u$ is
$x$-independent. The
reason for making such an assumption is that in most ballistic structures
the
scattering potential is rather soft. This is so because the
impurities are located
rather far from the region occupied by the conducting
electrons (see, e.g. Ref. \onlinecite{ando}).
Since we are interested in the role of an average potential, we may
neglect the weak spatial dependence of $\delta u$.
The  noise properties in our model are thus entirely due to the
dimensionless random quantity $\delta u(t)$.

The general expression for the current in the absence of intermode
scattering \cite{BeenHout} is,  %%PH: Better reference than: \cite{zsh1},
\begin{equation}
I(V_{sd},t)  =  \frac{2e}{h} \int dE \,  \delta   n_{F}(E) \,
\sum_{n} T_n[E, \delta u(t)],
\end{equation}
where
$T_n[E,\delta u(t)]$ is the probability of transmission
between the two
reservoirs for an electron in mode $n$
having energy $E$.  According to our simplified model, which will be used in
the our numerical calculations, we assume that  all  the  propagating
modes have the same transparency $T$ at the Fermi level. Thus,
\begin{equation}  \label{t_n}
T_n[E_{\text F}, \delta u(t)]=
{\bar T} \Theta[1-(n/\eta)^2 -\delta u(t)],
\end{equation}
where    $\Theta    (x)$    is   the   Heaviside   step   function.
The transmission probability is hence dependent on the quantity $\delta u(t)$.
The current is driven by the  difference in chemical potentials
of  the  Fermi-Dirac
distribution  functions  in the
reservoirs,
$$\delta n_{F}(E)  \equiv  n_{F}(E- E_{\text F} - eV_{sd}/2) -
	n_{F}(E- E_{\text F} + eV_{sd}/2).$$
(The total bias $V_{sd}$ is here chosen to be
symmetrically shared by the distributions
to the left and to the right.)
We have studied the low temperature regime ($T=0$) for small
source-drain voltages, where this difference simplifies to
$\delta n_{F}(E) =  eV_{sd} \delta(E- E_{\text F})$.
The current will in this case be,
\begin{equation}\label{curr}
I(V_{sd},t)  =  \frac{2e^2 V_{sd}}{h} \sum_{n} T_n[E_{\text F},\delta u(t)].
\end{equation}

Since no scattering between the electronic states
is taken into account, the only
mechanism for changing the current is the
possibility of having different numbers
of propagating modes at different times.

\section{Low-frequency noise } \label{noise}

\subsection{General consideration}

We consider fluctuations in $\delta u (t)$ as a random process
composed of contributions from many elementary fluctuators.
Each of them, let's say the $i$-th one,
is described by a normalized random variable $\xi_i(t)$
changing between the two values $\pm 1$  with
the corresponding transition rates $\Gamma_i^+$
and $\Gamma_i^-$.
Within the contact, this EF will create a potential of
strength $a_i$ for the electrons.
Transitions between the two different states are assumed to be
induced by interactions with a thermal bath.
The rates  $\Gamma_i^+$  and $\Gamma_i^-$ are in
general different and determined
by the nature of hopping and the interaction
between the  EF and the thermal bath.
{}From the detailed balance principle we find,
\begin{equation}  \label{Delta}
\Gamma_i^- /\Gamma_i^+ = \exp (-\Delta /k_{\text{B}}T),
\end{equation}
where  $T$  is  the  temperature and  $\Delta$  the energy difference
between the states of the EF.
For high enough temperatures the rates are almost equal, while
at low temperatures there will be a significant difference between the
two.
Here we consider the case $\Delta \ll k_BT\ll E_{\text F}$,
giving $\Gamma_i^-=\Gamma_i^+ \equiv \Gamma_i$.
The dependence of $\Gamma_i$ on $\Delta$ and $T$ is determined by the
hopping mechanism (see
discussion in Ref. \onlinecite{gkk}).
If the transitions are due to quantum mechanical tunneling the rates
are independent of temperature and we have
$\Gamma_i  \propto \Delta^k$. For EF-phonon interaction, $k=3$
while for EF-electron interaction, $k=1$.
If transitions  are  induced  by  activation,   $\Gamma_i \propto
\exp (-W/k_{\text{B}}T)$ where $W$ is some activation energy \cite{gkk}.

The number of fluctuators has been assumed to be large enough for
substituting
fixed values of the parameters for each fluctuator with values obtained from
continuous distributions over $a$ and $\Gamma$. We assume
these distributions to be uncorrelated,
${\cal P} (a, \Gamma) \equiv P_a(a)P_\Gamma(\Gamma)$.
The nature of the statistical process will be completely defined by
the distributions over $a$ and $\Gamma$ and the number of fluctuators (N).
Whether or not the noise
in practice is caused by internal or external parameters is actually
irrelevant. The result we have obtained applies to all cases when
electrons in the channel experience a time dependent potential
of the kind
$\delta u (t) = \sum_{i=1}^N a_i \xi_i(t)$,
where the distributions of the random quantities $a_i$ and $\Gamma_i$ are
to be discussed below.

The gate potential $\delta u(t)$ (see Sec.~\ref{curr_section}) is composed
of the electrostatic potential
from several EFs in the doping layer.
As an approximation we may nevertheless assume the EFs to be uniformly
distributed  in a spherical shell  $r_{\min} \leq r \leq r_{\max}$ ($r$
being the radial coordinate measured from the center of the channel)
in the doped region.
The probability for finding an EF in a shell between $r$ and  $r+dr$ is then
$P(r)dr \propto r^2dr$. If we subtract the static background charge,
we obtain a dipole potential, with a fluctuating sign
from each EF
%%PH why? variation
(see e.g. Ref.~\onlinecite{gkk}).
This implies that the interaction parameter $a$
will be proportional to the inverse of the cube of the separation,
$a \propto r^{-3}$.
Normalizing under the assumption $a_{\max}/a_{\min} \gg 1$,
we find the distribution of $P(a)$ to be
\begin{equation}\label{pa}
P_a(a) = \frac{a_{\min}}{a^2}.
\end{equation}
Even if the upper limit does not enter the expression for $P(a)$, it
is important to
include it in the following way. There must be an upper limit of
possible fluctuations
in the gate voltage since the number of fluctuators is finite. It will be given
by $Na_{\max}$. For larger gate voltages the one- and
two-point probabilities must
be zero. But the approximation we have adopted to find these
probabilities (see below)
is only asymptotically
($N \rightarrow \infty$) exact, and disregards this important fact.
Therefore we have neglected the finite value of
$a_{\max}$ throughout the derivation
but included it at the end by truncating the two
probability distributions over fluctuations in gate voltages.
Note that while $a_{\min}$ is related to the maximum distance to the
impurities, $a_{\max}$
is related to the minimum distance.

The transition probabilities $\Gamma^{\pm}$ usually have
an exponential dependence on the barrier parameters. Indeed,
assume the EF
to be described as an entity moving in a double well potential
\cite{and,phil,karp}. The logarithm of the transition probability
is then proportional either to the barrier height (in the case of thermal
activation), or to its strength (in the case of quantum tunneling).
Since these heights have been found
to have little spread among the different fluctuators, the
distribution of $\ln (\Gamma)$ is smooth and almost uniform.
We may hence model this distribution
to be constant in the region between $\Gamma_{\min}$ and
$\Gamma_{\max}$.
As the result we obtain
\begin{equation}\label{pgam}
P_\Gamma (\Gamma)= \frac{1}{\ln(\Gamma_{\max}/\Gamma_{\min})} \frac{1}{\Gamma},
\end{equation}
for the normalized distribution of $\Gamma$.

\subsection{Analytical expression for noise intensity}\label{exprnoise}

Noise is usually characterized by the current-current correlation function,
\begin{equation}  \label{ngen}
S(\tau) = \langle I(t+ \tau) I(t) \rangle_t - \langle I(t) \rangle_t^{2},
\end{equation}
or  by  its Fourier transform with respect to $\tau$, $S(\omega)$.
The symbol $\langle \cdots \rangle_t$ means average over $t$
which is (under stationary conditions) the same as an ensemble average
over the random process $\xi (t)$ (ergodicity).

{}From expression  (\ref{curr}) for the current,
we obtain the following
current-current correlation function,
\begin{eqnarray} \label{corrf}
S(\tau) &=& \left( \frac{2e^2 V_{sd}}{h} \right)^{2}
\sum_{n,m} \left[ \left< T_n(t) T_m(t+\tau) \right>_t
 \right. \nonumber \\ && - \left.  \left< T_n(t) \right>_t \left<
T_m(t) \right>_t \right].
\end{eqnarray}
The transmittance is $T_n(t) \equiv T_n[E_{\text F}, \delta u (t)]$.
In order to simplify the notation,
we from now on substitute $\delta u$ with $u$.
Since the gate voltage fluctuations must
become uncorrelated for large
time differences, expression (\ref{corrf}) may be written as
\begin{eqnarray}
S(\tau) & = & \left( \frac{2e^2 V_{sd}}{h} \right)^{2}
\sum_{n,m} \left[ G(n,m | \tau) - G(n,m | \infty) \right] \, ,  \\
G(n,m | \tau) & = & \left< T_n(t) T_m(t+\tau) \right>_t
\nonumber \\
&& = \int du \int dv T_n(u) T_m(v) P_2(u,v | \tau).
\end{eqnarray}
The two-point probability of having gate voltage
fluctuations
$u$ and $v$ at times differing by $\tau$ is $P_2(u,v | \tau)$.
$P_2$ must split into one-point
probabilities $P_1$ for large values of $\tau$,
$\lim_{\tau \rightarrow \infty}  P_2(u,v | \tau) = P_1(u) P_1(v)$.

The probabilities $P_1$ and $P_2$ may be calculated using a generating
function,
\begin{equation}
K_N (x,y)= \left< e^{-ixu(t)-iyv(t+\tau)} \right>_{\xi_i},
\end{equation}
where $N$ denotes the number of fluctuators.
The average is over all random quantities $\xi_i(t)$ representing
elementary fluctuator $i$ with strength $a_i$ and transition
rate $\Gamma_i$.
{}From the definition of expectation values,
\begin{eqnarray}
K_N(x,y|\tau) & = & \int e^{-ixu-iyv}P_2(u,v|\tau)du dv, \\
K_N(x,0|0) & = & \int e^{-ixu}P_1(u)du.
\end{eqnarray}
These relations can easily be inverted to find $P_1$ and $P_2$ as
functions of $K_N$,
\begin{eqnarray}
P_2(u,v|\tau) & = & \frac{1}{(2 \pi)^2} \int e^{ixu+iyv}
K_N(x,y|\tau) dx dy  \label{P2}, \\
P_1(u) & = & \frac{1}{2 \pi} \int e^{ixu} K_N(x,0|0) dx \label{P1}.
\end{eqnarray}

Provided the elementary fluctuators are statistically independent,
$K_N$ transforms
into a product,
\begin{equation}
K_N (x,y|\tau)= \prod_{i=1}^{N} \left< e^{-ixa_i \xi_i(t)-iya_i \xi_i (t+\tau)}
	\right>_{\xi_i}.
\end{equation}
Now consider factor number $i$ only.
This is a strictly one-fluctuator quantity, and
instead of fixing the values of the constants $a_i$ and $\Gamma_i$ we may
introduce distribution functions over $a$ and $\Gamma$.
Since then we have a function of these two parameters
we have to perform a weighting over
these distributions as well as the average over the statistical variable $\xi$,
\begin{equation}
\langle K_N (x,y|\tau)\rangle_e =  \langle K_1(x,y|\tau) \rangle_e^N,
\label{Kdef}\end{equation}
where $\langle \cdots \rangle_e$ means an average over
parameters $a, \Gamma$ describing the environment, while
\begin{equation}
K_1(x,y|\tau) =  \left< e^{-ixa \xi(t)-iya \xi (t+\tau)}
	\right>_{\xi}.
\label{K1}
\end{equation}
The last expression depends on the parameters $a$ and $\Gamma $.
In order to evaluate $K_N$ we apply the Holtsmark procedure
 \cite{holt},
which is applicable in the limit of many fluctuators,
\begin{eqnarray}
K_N (x,y|\tau) & = & e^{-F_N(x,y|\tau)}, \label{kn}
\nonumber \\
F_N(x,y|\tau) &=& N[1-\langle K_1 (x,y|\tau) \rangle_e].
 \label{fn}
\end{eqnarray}
This approximation is valid provided $1-K_1(x,y|\tau) \ll 1$ and $N \gg 1$.
The generating function $K_1(x,y|\tau)$ for one fluctuator and
fixed values of $a$ and $\Gamma$
has been calculated earlier \cite{rytov} (see also Refs.
\onlinecite{ggk1,hessgalp1}),
\begin{eqnarray}\label{K}
K_1(x,y|\tau)&=& e^{-\Gamma |\tau|} \left[ \cosh(\Gamma
|\tau|)\cos\left(a(x+y)\right) \right. \nonumber \\
	&&+ \left. \sinh(\Gamma |\tau|)\cos\left(a(x-y)\right)\right].
\end{eqnarray}
As is clearly seen, this expression decouples for large times
into a product of functions of $x$ and $y$
as it must if gate voltage fluctuations at large time
differences are uncorrelated.
%%The definition of $K_1$, (\ref{Kdef}) together with
Equation (\ref{K}) implies
that $F_N$ in Eq. (\ref{fn}) will be given by
\begin{eqnarray}
F_N(x,y|\tau)
%       &=& N(K_1(x,y|\tau)-1) \\
%	& = & N \int P_a(a)
%	\frac{\cos[a(x+y)] + \cos[a(x-y)]}{2} da \\
%	& + &
%	N \int P_\Gamma(\Gamma) e^{-2\Gamma|\tau|} d\Gamma
%	\int P_a(a) \frac{\cos[a(x+y)]-\cos[a(x-y)]}{2} da \\
	& \equiv & -N\left\{ \Phi(x+y) + \Phi(x-y)
\right. \nonumber \\ && +\left.  \Psi(\tau)
	\left[\Phi(x+y)-\Phi(x-y)\right] \right\}.
\end{eqnarray}
The functions $\Phi$ and $\Psi$ are here defined as
\begin{eqnarray}
\Phi(x) & = & \frac{1}{2} \int_{a_{\min}}^{a_{\max}} P_a(a)
[1-\cos(ax)] \, da \nonumber  \\
&=& \frac{a_{\min}}{2} \int_{a_{\min}}^{a_{\max}} \frac{
1-\cos(ax)}{a^2} da \, ,  \label{Phi1} \\
\Psi(\tau) & = & \int_{\Gamma_{min}}^{\Gamma_{max}}
P_\Gamma(\Gamma) e^{-2\Gamma|\tau|} d\Gamma. \label{Psi}
% This is a good definition of psi.
\end{eqnarray}
Here we have made use of the distribution over fluctuator strengths (\ref{pa}).
Using (\ref{pgam}),
$\Psi(\tau)$ can only be calculated analytically in the asymptotic limits,
\begin{equation}\label{psiasym}
\Psi(\tau) = \left\{
\begin{array}{cl}
1- \frac{2|\tau|\Gamma_{\max}}{\ln(\Gamma_{\max}/\Gamma_{\min})},
 & |\tau| \ll \Gamma_{\max}^{-1} \\
\frac {\ln(1/|\tau|\Gamma_{\min})}{\ln(\Gamma_{\max}/\Gamma_{\min})}, &
	\Gamma_{\max}^{-1} \ll |\tau| \ll \Gamma_{\min}^{-1} \\
0, & |\tau| \gg \Gamma_{\min}^{-1}
\end{array}
\right..
\end{equation}
The product  $Na_{\min}$,  which  in  the  following  will  be  denoted  as
$4C_N/\pi$, has the physical meaning of a typical shift of the QPC potential.
We will  show below  that  the typical value of $x$ in the
function $\Phi (x)$ is $C_N^{-1}$. Hence, we replace the  lower
limit  of  integration in Eq. (\ref{Phi1}) by 0. The behavior of $\Phi (x)$
depends  on  the relation between the upper limit $a_{\max}$ and $C_N$.
We can also replace the upper limit by infinity provided,
\begin{equation}  \label{maxa}
a_{\max} \gg C_N, \ \text {or} \ r_{\min} \ll {\bar r}.
\end{equation}
The constants $r_{\min}$ and $\bar r$ are here the minimum and average
distance, respectively,  between the QPC and the EF.
Taking the derivative of $\Phi$ twice, we obtain an integral representation
of the Dirac delta function.
Using its properties while integrating repeatedly we find,
\begin{equation}
\Phi(x)= \frac{\pi a_{\min}}{4} |x|, \quad a_{\max} x \gg 1.
\end{equation}
In the opposite limiting case, %%%%%%%to (\ref{maxa}),
\begin{equation}  \label{phig}
\Phi (x) = \frac {a_{\min}a_{\max}}{4}x^2, \quad a_{\max} x \ll 1.
\end{equation}
Under the conditions (\ref{maxa}) $F_N$ in Eq. \ref{fn} is found to be
\begin{eqnarray} \label{fxy}
F_N(x,y,\tau)&=& -C_N[(1+\Psi(\tau))|x+y|
\nonumber \\ && + (1-\Psi(\tau))|x-y|].
\end{eqnarray}
The Fourier transform with respect to $x$ and $y$
needed to get $P_2$, given by Eq. (\ref{P2}), is readily
found from (\ref{kn}) and (\ref{fxy}) to be a
product of two Lorentzian functions.
Similarly, from (\ref{P1}) we calculate $P_1$
to be a simple Lorentzian function.
Unfortunately, because of the slow decay of the Lorentzians,
our approximations made so far are not sufficient for  calculating
the noise. We must also take into
account that for large values  of the fluctuations
($u,v \ge  Na_{\max}$),  the  approximation
(\ref{fxy})  breaks down.
%We  do  this  by  introduction  of  the
%truncated Lorentzian function
%\begin{equation}  \label{callor }
%{\cal L}(z,w) =\frac {\Theta (Na_{\max}-|z|)}{2\arctan (Na_{\max}/w)}
%\frac {w}{z^2+w^2}.
%\end{equation}
We correct for this by introducing truncation factors ${\cal D}$
for each variable representing fluctuations in the
gate voltage in the probabilities,
\begin{eqnarray}
P_1(u) & = & {\cal D}_1(u){\cal L}(u|2C_N) \, ,\label{P1u} \\
P_2(u,v|\Psi) & = &\frac{1}{2} {\cal D}_2(u) {\cal D}_2(v)
{\cal L}_+ {\cal L}_-,
\label{P2uv} \\
{\cal L}_+ &=&{\cal L} \left(\left.  \frac{u+v}{2}   \right|
C_N(1+\Psi)\right),
\\
{\cal L}_- &=&
{\cal L}\left(\left.\frac{u-v}{2} \right| C_N(1-\Psi)\right), \\
{\cal D}_i(u) & = & \frac{\Theta (Na_{\max}-|u|)}{{\cal N}_i}, \\
{\cal L}(z,w) & = & \frac{1}{\pi}\frac{w}{z^2+w^2}.
\end{eqnarray}
The function $\Psi(\tau)$ is given by (\ref{Psi}),
and has the asymptotic values  1
for $t \rightarrow 0 $ and 0 for $t \rightarrow \infty$.
The necessary renormalization factor imposed by the truncation procedure
is ${\cal N}_i$. For $P_2$
this constant has to be calculated numerically, but for $P_1$,
${\cal N}_1=2\arctan({Na_{max}/\pi})$.
Note  that  such  a  truncation  has  an  obvious  physical  meaning -- the
fluctuation of, say, $u$ cannot be greater than %%%the maximal potential shift
$Na_{\max}$.
This shift represents the case when all fluctuator contributions
add up coherently to maximal total potential felt by the electrons.

%Collecting, we hence arrive at the following expression for the
%current-current correlation function,
%\begin{eqnarray}\label{curr-corr}
%S(\Psi(\tau)) & = & \left( \frac{e^2 V_{sd}T}{ \pi \hbar} \right)^{2}
%\sum_{n,m} \int_{-\infty}^{w_n} du \int_{-\infty}^{w_m} dv
%\left[ P_2(v,u | \Psi) - P_2(v,u |0) \right] \\
%P_2(v,u|\Psi) & = & {\cal D}(u) {\cal D}(v)
%\frac{1}{2} {\cal L} \left(\left.  \frac{u+v}{2}   \right|
%C_N(1+\Psi)\right)
%{\cal L}\left(\left.\frac{u-v}{2} \right| C_N(1-\Psi)\right) . \label{p2}
%\end{eqnarray}
%where $\Psi(\tau)$ is given by (\ref{Psi}), and has the asymptotic value  1
%at $t \rightarrow $ and 0 at $t \rightarrow \infty$.

The final expression for the current-current correlation function will be
\begin{eqnarray}\label{curr-corr}
S(\tau) &=& \left( \frac{2e^2 V_{sd}}{h} \right)^{2}
\sum_{n,m} \int_{-\infty}^{w_n} du \int_{-\infty}^{w_m} dv
\nonumber \\ && \times
\left[ P_2(u,v | \Psi (\tau)) - P_2(u,v |0) \right].
\end{eqnarray}
The upper limits of integration are here given by $w_n \equiv 1- (n/\eta)^2$,
and $P_2$ by (\ref{P2uv}) assuming
${\bar T}=1$.
Finally, the noise spectrum is found as the Fourier transform of $S(\tau)$.

Let us compare our result with the conventional formalism
for calculating noise in electronic systems.
To be general, assume
the current to be dependent on a random quantity $u$ that fluctuates in time
around the average value $u_0=\langle u \rangle_t$.
In our case it will be the effective  gate voltage but the nature of
this quantity is not of crucial importance here.
As a random variable, $u$ possesses certain one- and two-point probabilities
$P_1(u)$ and $P_2(u,v|\tau)$, similar to what was presented in the
preceding section. Further, consider the case for which
the current is a smooth function of $u$ and
typical fluctuations are small.
Then we are allowed to
expand the current as a function of the variable $u$,
\begin{equation}\label{Iexp}
I[u(t)]= I(u_0)+ G(u_0) \delta u(t) + {\cal O}\{[u(t)]^2\},
\end{equation}
where the transconductance $G$ is defined as $G (u_0) =\partial I /
\partial u \vert|_{u=u_o}$.
Note that in our case this should {\em not} at all be an adequate procedure --
the transmission probability being one for propagating modes and zero for
reflected ones is certainly not a differentiable function of the
gate voltage.
Nevertheless we may compare the two results
to see if they deviate significantly.
The expansion
Eq. (\ref{Iexp}) directly implies that an approximate
expression for the current-current
correlation function can be written down as
\begin{eqnarray}\label{Strc}
S(\tau) &=& \langle I(t+ \tau) I(t) \rangle_t - \langle I(t) \rangle_t^{2}
\nonumber \\
&&\approx G^2(u_0) \langle \delta u(t+ \tau) \delta u(t) \rangle_t .
\end{eqnarray}
In this approach the correlation function factorizes into two
parts. One  describes the rigid part of the system and the other
is solely determined by the time-dependent random quantities.
As previously,
substitute $\delta u$ with $u$ to simplify the notation.

In our study, $u$ is the fluctuating part of the gate voltage and
$G$ the transconductance. The latter can be evaluated as the
derivative of the average current with respect to the gate voltage.
{}From (\ref{curr}) and (\ref{P1u}) we easily find the average current to be,
\begin{eqnarray}
\langle I(t) \rangle_t & = & \frac{2e^2 V_{sd}T}{h {\cal N}}\sum_{k=1}
\left[ \frac{1}{\pi}\arctan\left(\frac{u_k}{2C_N}\right)+{\cal N}/2 \right], \\
u_k & = & \min \left\{ {1-\left(\frac{k}{\eta}\right)^2,Na_{max}} \right\},
\end{eqnarray}
where the sum is over propagating modes.
The transconductance will then be,
\begin{equation}\label{trc}
G(\eta)= -\frac{2e^2 V_{sd}T}{h {\cal N}}\sum_{k=k_{\min}}^{k_{\max}}
{\cal L}\left(\left. 1-\left(\frac{k}{\eta}\right)^2 \right| 2C_N\right).
\end{equation}
Here, $k_{\min}= [1+\eta\sqrt{1-Na_{\max}}]$ and $k_{\max}=
[\eta\sqrt{1+Na_{\max}}]$;
$[ \cdots]$ denotes the integer part.
$\eta$ is related to $u_0$ according to (\ref{n0}).
The average
 $\langle u(t+ \tau) u(t) \rangle_t$  can be evaluated
when the truncation is done in the variables $(u \pm v)/2$ instead of
$u$ and $v$.
This modification is not less accurate than the truncation itself.
Making use of (\ref{P2uv}) but with this modified truncation we find,
\begin{eqnarray}\label{uv}
&&\langle u(t+ \tau) u(t) \rangle_t  =
-4 \Psi C_N^2
\nonumber \\
&&+ (Na_{\max})^2\left[{\cal
A}\left(\frac{B}{1+\Psi(\tau)}\right)-
	{\cal A}\left(\frac{B}{1-\Psi(\tau)}\right)\right],
\label{uvave} \\
&&{\cal A}(x)  = 1/x \arctan(x), \quad
B = 4a_{\max}/\pi a_{\min},
\end{eqnarray}
where $\Psi(\tau)$ is given by (\ref{Psi}).
If $Na_{\max}\gg \pi C_N$ (which is also required for using the
Holtsmark approximation
\cite{holt}), the obvious restriction
$\lim_{\tau \rightarrow 0} \langle u(t+ \tau) u(t) \rangle_t > 0$
is fulfilled.
Furthermore, since $\langle u(t) \rangle_t = 0$ we must have
$\lim_{\tau \rightarrow \infty} \langle u(t+ \tau) u(t) \rangle_t =
\langle u(t+\tau) \rangle_t \langle  u(t) \rangle_t = 0$,
where we have assumed that fluctuations in the effective gate voltage
become uncorrelated at  large time differences.
Obviously, (\ref{uvave}) obeys this condition.
Collecting Eqs. (\ref{Strc}), (\ref{trc}) and (\ref{uv}) we may compare this
conventional approach of evaluating noise
to our calculation. In Fig.~(\ref{fig:1})
we indeed find, as expected,
significant deviations between the two treatments.
At the second conductance step,
they differ approximately by a factor $25$. It should be
emphasized that this
comparison is strongly dependent on the choice of the parameter $Na_{\min}$.

The case opposite to (\ref{maxa}) for which $a_{\max} \geq a_{\min}$,
is applicable to a structure with
a very wide
spacer compared to the thickness of the doping layer.
%%%%PH Why?: and high level of doping.
The Lorentzians ${\cal L}(z|w)$ in
Eq. (\ref{P1u}) - (\ref{P2uv})
are then replaced by  Gaussians,
$${\cal G}(z|w)=\frac {1}{w\sqrt{2\pi}}e^{-z^2/2w^2},$$
and the quantity $C_N$ by $N\sqrt{a_{\min}a_{\max}}$ [see
Eq. (\ref{phig})].  However, we  do  not further pursue the analysis of
this  situation.

\subsection{Numerical analysis}

We had to resort to a numerical calculation of the noise using a discrete
cosine transform.

In general there is a problem with
discrete transforms that is called aliasing
\cite{numrec}.
It occurs when the time sequence is
not bandwidth limited, and shows up as an increased Fourier transform
for high frequencies. In short, high frequency components
outside the range where
the transform can be calculated are folded into this domain.
For the case of investigating $1/f$-like noise
this problem is severe.
The contribution to the transform that is
transferred into the region of interest
would even diverge if the transform was exactly proportional to $1/f$.
When the signal has a finite bandwidth this problem does not occur.
To invent an artificial bandwidth in our case we decrease
the physical parameter
$\Gamma_{\max}$ to be of the same
order as the upper limit of the Fourier transform of $S(\tau)$.
Evidently, this will be allowed since the numerically calculated transform
anyway will be cut in the vicinity of the upper limit.

If the function of time can only be sampled in a relatively few number
of points because of computational difficulties there is also a problem
to find the transform over a wide range of frequencies. Basically,
it is only possible to find the transform over a number of decades
of the order $\left[\log_{10}(\mbox{number of sample points})-1\right]$.
We solve this problem in the following
way.
The time scale is set by the upper limit of $\tau$, $\tau_{\max}$.
By changing its value we may successively map out different parts
of the Fourier transform over a wide range of frequencies.
For low frequencies ($\omega<100 \Gamma_{\min})$, a large value of
$\tau_{\max}$ is used and $S(\tau)$ is sampled over almost the entire
domain.

For high frequencies though, only the most rapidly varying part of
$S(\tau)$ is relevant because the contribution from the remaining parts
is averaged out when the transform is calculated. Hence it is sufficient to
select the most rapid variations, which in our case are close to $\tau=0$.
Keeping the same number of sample points while $\tau_{\max}$ is decreased,
the resolution in time is
enhanced. With an increased resolution we are able to compute
the high frequency components. Note also that when changing $\tau_{\max}$
we must change $\Gamma_{\max}$ accordingly in the way mentioned above.
Unfortunately, because we don't sample $S(\tau)$ for all values of $\tau$
two problems occur. Firstly, the calculated transform will have an offset.
We correct this by adjusting the transform to the low frequency solution
where $\tau_{\max}$ is very large. Secondly, there will be rapid oscillations
in the transform because we truncate $S(\tau)$ with a sharp step when
$\tau_{\max}$ is decreased.
This problem is solved by filtering, averaging over a small interval
yields the correct form of the transform.
In practice we convolve
the Fourier transform using a constant filtering function with
sufficiently large bandwidth (in time).
(Note that time and frequency are exchanged here compared to the
'normal' case when we filter a function of time.)
We repeat the calculation and decrease $\tau_{\max}$ in
sufficiently small steps (we used a factor of $10$) every time.
Finally, we
put the pieces together and match each new section at the
lowest frequency.
%
%%%%%%%%%%% PH: previous junk
%
%Then we have computed the transform for different values of the upper limit in
%$\tau$, $\tau_{\max}$, with the same number of points, using
%different values of $\Gamma_{\max}$.
%For mapping out the high frequency components, a rather small value of
%$\tau_{\max}$ has to be used.
%Because in this way we select the most rapidly varying
%part of the time sequence
%(in our case for small values of $\tau$) which has the strongest influence
%for high frequencies.
%However, this cut off will introduce oscillations
%in the transform as any sharp truncation would, but will also cause the
%numerically calculated transform
%to be a fractional part of the true transform.
%The latter problem is solved by using a sufficiently large value of
%$\tau_{\max}$ for the lowest frequencies ($\omega<100 \Gamma_{\min}$),
%for which no filtering is done.
%Then we decrease $\tau_{\max}$ in several steps
%(by a factor of $10$ in each)
%to reach higher and higher frequencies and adjust the off set value of
%$\log [S(\omega)]$ after filtering to the lowest frequencies.
%We have not found any general motivation why this procedure should work
%in principle,
%but if the pieces matches in a very regular way, the result at least looks
%plausible.
%
In this way we are able to find the transform over a
wide frequency range.
Under any circumstances, the transform
for $\omega<100 \Gamma_{\min}$ is not affected at all by our method of
calculation.
A smooth matching (see Fig.~\ref{fig:2}) of the pieces
provides some numerical evidence that the method
is adequate.

The parameter $(Na_{\min}/E_{\text F})$
describing a typical fluctuation in the potential from all impurities
was chosen to be equal to $5\cdot 10^{-3}$;
the maximal fluctuation was determined by $a_{\max}/a_{\min}=100$.
For the Holtsmark approximation \cite{holt} to be valid a large
ratio between the maximal and average potential fluctuation is required,
$a_{\max}/Na_{\min} \gg 1$
(see also the end of subsection \ref{exprnoise}).
With these parameters we may for instance
have $10$ impurities implying $a_{\max}/Na_{\min}=10$, which seems
plausible.
As has already been mentioned,
the transmission probability $T_n(u)$ was taken to
be zero for non conducting modes and unity for conducting ones.
In this case we will find that $\sum_{n=1}^{\infty} T_n(u) = [\eta]$,
where $[ \eta ]$ denotes the integer part of $\eta$,
which gives the number of conducting modes
at zero fluctuation in gate voltage.

%%%%%%%%%%%%%%%%% FIG. 1.EPS %%%%%%%%%%%%%%%%%%%%%

A comparison of our calculation with the simple-minded
conventional approach leading to (\ref{Strc}) is done in Fig.~\ref{fig:1}.
Large deviations are found for this set of parameters,
at the step $S_B(\omega)/S_A(\omega) \approx 25$.
Indeed, significant deviations are expected due to the
non-differentiable character of the transmittance as a
function of gate voltage. The step-like shape of the average
transmittance is responsible for the strongly enhanced
level at the step and the very low level of noise on the
plateaus.

%%%%%%%%%%%%%%%% FIG. 2.EPS %%%%%%%%%%%%%%%%%%%%%%

In Fig.~\ref{fig:2}, the logarithm of the noise is plotted
in units of $(2e^2 V_{sd}/h)^2$
versus the logarithm of
dimensionless frequency $w= \omega/\Gamma_{\min}$.
The graphs show the noise
at the second conductance step ($\eta=2$) as
well as between the second and third step ($\eta=2.5$).
Generally, the current fluctuations are maximal
when $\eta$ is an integer.
This corresponds to a step in the current-gate voltage curve.
Since noise is just the Fourier transform
of the current-current correlation function, we consequently expect
the noise to be large for integer and small for non integer
values of $\eta$. This is also clearly seen in Fig.~\ref{fig:2}.

If $Na_{\max} \leq E_{\text F}/[\eta]$
noise is actually independent of $a_{\min}$ {\em at} the step.
This can be understood in the
following way. The two Lorentzian functions
in the two point probability (\ref{P2uv})
is in this case split
symmetrically into two regions of positive and negative gate voltage
fluctuations.
These two domains are
defined by $T_n$ being constant.
The tails of the probability distribution only extend into
the nearest plateau on both sides of
the step for this value of $a_{\max}$.
It is then irrelevant how wide the Lorentzians
are (determined by $Na_{\min}/E_{\text F}$),
as long as the total weight of probability
remains the same in the different
domains.
Only the probability density {\em profile} differ
for different values of
$Na_{\min}/E_{\text F}$. But since $T_n$ is
constant in both regions, it has no effect on $S$.
By measuring the level of noise at and
between the steps we may infer the order of magnitude of the parameter
$a_{\min}$. It will be related to the ratio of the
two noise intensities provided $Na_{\max} \leq E_{\text F}/[\eta]$.
For the presented results in Fig.~\ref{fig:2}, we actually have
$Na_{\max}=E_{\text F}/[\eta]$.
%It has been numerically verified that the
%same result was obtained for $\eta=2$ if $Na_{\max}$ was decreased
%to $0.2E_{\text F}/[\eta]$.

Between the steps in the current-gate voltage characteristics
the noise must decrease with $a_{\min}$, since
the average potential fluctuations are proportional to this parameter.
%, which is also seen in Fig. (\ref{fig:2}).
This has also been checked.
%%%%%%%% PH: No point in having it here (below)
%Crudely speaking,
%the time dependence in $S(\tau)$ and $\Psi(\tau)$ is similar.
%For the maximum frequency in our calculation,
%$10^5\Gamma_{\min}$, only $S(\tau)$ for times
%$\tau$ larger
%than $10^7\Gamma_{\min}^{-1}$ and smaller than
%$\Gamma_{\min}^{-1}$ is important (except for $\tau=0$).
%In  this regime one may infer the qualitative
%behavior of $S(\tau)$ from the approximate form of $\Psi$,
%(\ref{psiasym}).

{}From Fig.~\ref{fig:2} we deduce that in all cases,
the dependence is almost linear for frequencies
larger than some typical value %%%%we may denote
$\omega_c$. It is determined by the parameters
$\eta$ and $Na_{\min}/E_{\text F}$.
For $\eta=2$ corresponding to the second step in conductance
quantization,
$\omega_c \approx 2\Gamma_{\min}$ irrespective of the value of
$Na_{\min}/E_{\text F}$
when this quotient is sufficiently small ($<1 \cdot 10^{-3}$).
If the system is biased between
the second and third step $\eta=2.5$, $\omega_c$
is of the order $10 \Gamma_{\min}$.
This threshold frequency has been found to slightly increase
with decreasing $Na_{\min}$ between the steps.
%\footnote{Check}
Because of computational difficulties,
the form of the curves for, let's say $\omega > 10^4 \Gamma_{\min}$,
should only be considered to be
accurate within a relative error of the order $5\%$.
The slopes for $\omega > \omega_c$ were found to be close to $ -0.85$
($-0.93$) for $\eta = 2.0$ ($\eta = 2.5$),
$S(\omega) \propto 1/\omega^{0.85 (0.93)}$.
This means that the noise is almost
$1/f$-like.

%%%%%%%%%%%%%%%% FIG. 3.EPS %%%%%%%%%%%%%%%%%

Experimentally, one is also interested in
the integrated noise intensity per decade.
For the curves in Fig.~\ref{fig:2},
this quantity is shown in Fig.~\ref{fig:3}.
The integration is for decade number $k$
performed over the interval $[k-0.5,k+0.5]$.
The center frequency is here $\omega_c$, which in turn defines
$k$, $k=\omega_c/\Gamma_{\min}$.
If the noise was identically $1/f$,
the integrated noise intensity per decade would be constant
for all decades.
%Any portions in the noise with a negative
%(positive) second derivative would imply
%the integrated noise to increase (decrease)
%with the decade number.
In reality, the noise must
be cut for frequencies larger than the physical value
of $\Gamma_{\max}$. In realistic systems $\Gamma_{\max}$
can be as large as
$10^{12}\Gamma_{\min}$.
This is beyond the limits of our calculation,
so we are unable to detect this physical cut-off.

\section{Discussion and conclusions}\label{discussion}

We would like to stress that the calculated level of noise
significantly differed
from a conventional estimate, see Fig.~\ref{fig:1}.
(It is lower by a factor $25$ for one set of parameters).
The approximation was
based upon an expansion of
the current-current correlation function in fluctuations
of the gate voltage.
Due to the sharp structure of the average transmittance
$\langle T_n \rangle_t$ the conventional approach leads
to invalid results. The deviation from the true result
depends on the strength of the fluctuators --
the weaker they are the larger is the difference.

According to our model the current-current correlation
function $S(\tau)$ (\ref{curr-corr})
has the following properties.
Under certain restrictions (see Appendix \ref{app:samenoise})
it is independent of the number of
propagating modes, $[\eta]$. Furthermore, if the spacer is thick enough  and
the inequality  $Na_{\max} < E_{\text F}/[\eta]$ holds,
$S(\tau)$ is also independent of $a_{\min}$
at the steps in conductance quantization.
Hence, measuring the noise level at a step would be enough to
find the value of $a_{\max}$ which is the interaction strength between
the nearest elementary fluctuator and the QPC.

 %the framework of
Our model (which assumes a constant transmittance $T_n$
on every plateau)
suggests that the level of noise is very low close to the
middle of a plateau.
Generally, it is more dependent on  the maximum
($r_{\max} \propto 1/a_{\min}$) than the
minimum ($r_{\min} \propto 1/a_{\max}$) distance to the EFs.
This fact provides us with a way of determining
$a_{\min}$ almost independently of $a_{\max}$. Since $a_{\min}$ then
effectively is the only parameter, the level of noise at a plateau where
many modes are conducting should provide
sufficient information to determine its value. Having obtained both
parameters of the model one can examine its validity by comparing the
predicted and measured dependence of the noise on the gate voltage.

Numerical calculations indicate that there
is a typical frequency, $\omega_c$, specific for
each set of parameters and
above which $S(\omega) \propto 1/\omega^\alpha$.
The exponent
$\alpha$ was found to be slightly less than $1$ and
the threshold frequency $\omega_c$ close to
the minimal transition rate $\Gamma_{\min}$ for the fluctuators.
Also, on the plateaus $\alpha$ is slightly enhanced compared to
its value at the steps.
Further, $\omega_c$ increased when the minimum fluctuator
strength $a_{\min}$ decreased.

Experiments on $1/f$-noise in QPC  \cite{yuan}
are in qualitative agreement with our theoretical results.
%\footnote{Temperature 4.2K low/high compared to $E_{\text F}$
% ?? maybe add discussion - Peter}
Indeed, the noise was
found to be $S(\omega) \propto I^{\beta}/\omega^{\alpha} $,
where $\beta=2\pm0.4$ and $\alpha=0.9\pm0.1$.
As we also conclude, the value of $\alpha$ was larger on the plateaus
than at the steps
in conductance quantization with gate voltage.
Of course, the level of the noise was larger at the steps than between them.
For the highest two values of the number of conducting modes,
$4$ and $5$, the noise was almost the same
in agreement with our claims (see appendix \ref{app:samenoise}).
It was also pointed out that after some cool-downs of the sample
the peaks of the noise as a function of the parameter $\eta$
were smeared out. As a possible cause for this effect we
suggest that the elementary fluctuators may
change their positions between subsequent measurements because of heating.
This would affect $a_{\min}$ as well as $a_{\max}$ and
give rise to the observed effect.

Finally, we recall that our derivation of noise is done in the limit
of many fluctuators, $N \gg 1$. At the same time, we assume that
$N \ll (r_{\max}/r_{\min})^2$. The last inequality means that a
significant number of EFs are situated in a close vicinity of the QPC.
The probability to have a large time-dependent fluctuation of the
electric potential is large and governed by
a Lorentzian distribution function (\ref{P2uv}). If the last
inequality does not hold only the nearest EF is of importance and the
distribution is a Gaussian function. If $N$  is of order 1, the
noise has a random telegraph character.

\section{Acknowledgments}

This work was supported by NorFA,
grant no. 93.30.155 and by the Nordic Research Network on the Physics of
Nanometer Electronic Devices.
JPH gratefully acknowledges the hospitality
of the Department of Physics at Oslo
University, Norway.
We would like to thank Mats Jonson, Alexander Zagoskin and Robert Shekhter
for proofreading the manuscript.

\appendix
\section{Noise for different numbers of propagating modes}\label{app:samenoise}
The magnitude of the noise will here be shown
to be independent, under certain restrictions,
of the number of
propagating modes through the contact.
Let
zero fluctuation in gate
voltage correspond to a fixed position on the
quantized plateaus in the current-gate voltage
characteristics.
Assume that the Lorentzians in the two-point probability
extend over
%over intervals
%in the gate voltage shorter than or equal to a
%length corresponding to
$p$ steps in the
$I-V_g$ curve.
This provides a restriction on the value of $\eta$, $\eta \geq p$.
We have then $T_n(u)=1$ for all
modes $n \leq [\eta]-p \equiv q$
and gate voltages $u$ and $v$
for which $P_2(u,v|\Psi(\tau))=P_2(u,v|\tau)$ is finite.
The dimensionless correlation function
$s(\tau) \equiv (h/2e^2V_{sd})^2 S(\tau)$
 (see \ref{curr-corr}) can then be reduced,
\begin{equation}
s(\tau) =
 \sum_{n,m=q+1}
%\sum_{m=q+1}
 \int du \int dv \, T_m(u) T_n(v)
{\cal P}(u,v|\tau)
\end{equation}
where ${\cal P}(u,v|\tau)= P_2(u,v | \tau)- P_2(u,v |0)$.
Indeed, for terms with both $T$s constant, $T_n=T_m \equiv 1$
the integral
$\int du \int dv {\cal P}(u,v|\tau)$ vanishes because of
the normalization
condition.
On the other hand, if only one $T$ is constant, let's say $T_m \equiv 1$
the integrals
$\int du {\cal P}(u,v|\tau)$ vanish because
$$\int du P_2 (u,v|\tau) = P_1(v)$$
is $\tau$-independent.
%
%\begin{eqnarray}
%S(\tau) \left( \frac{ \pi \hbar}{e^2 V_{sd}} \right)^{2}&  =&
%\sum_{n=0} \sum_{m=0}  \int du \int dv \, T_n(u) T_m(v)
%\delta P_2(v,u | \tau) \nonumber \\
%& = & \sum_{n=0}^{q} \sum_{m=0}^{q} \int du \int dv \,
%\delta P_2(v,u | \tau)
% +  \sum_{n=q+1} \sum_{m=0}^{q} \int du \, T_n(u)
%\int dv \, \delta P_2(v,u | \tau)
% +  \sum_{n=0}^{q} \sum_{m=q+1}  \int dv \, T_m(v)
% \int du \, \delta P_2(v,u | \tau) \nonumber \\
%&& +  \sum_{n=q+1} \sum_{m=q+1}  \int dv \int du \, T_m(v) T_n(u)
% \int du \, \delta P_2(v,u | \tau) \nonumber \\
%& =&
%\end{eqnarray}
%
%
%The first term is zero because $P_2$ is normalized
%to one while the second and third term
%vanishes since,
%\begin{equation}
%\int du \, P_2(v,u | \tau) = \int du \,
%P_1(v)Q(u,v|\tau) = P_1(v) \int du \, Q(u,v|\tau) = P_1(v)
%\end{equation}
%which is independent of $\tau$.
%($Q(u,v|\tau)$ is the conditional probability of having gate
%voltage $u$ at time $t+\tau$ given $v$ at time $t$.)
Note that $T_n$ is dependent of the offset
gate voltage which is related to $\eta$.
Because of the simple form of $T_n$,
being zero for reflected modes and one for propagating
modes, we are able to make a translation in the mode index,
\begin{equation}
T_n(u,\eta)= T_{n-q}(u,\eta-q).
\end{equation}
Finally, changing indices of summation we obtain,
\begin{eqnarray}
s(\tau) &=&
%\left( \frac{e^2 V_{sd}}{ \pi \hbar} \right)^{2}
\sum_{n^\prime,m^\prime=1} \int du \int dv \nonumber \\
&&\times
T_{m^\prime}(u,\{\eta\}+p) T_{n^\prime}(v,\{\eta\}+p)
{\cal P}(u,v | \tau)
\end{eqnarray}
where $\{\eta\} \equiv \eta-[\eta]$ is the fractional part of $\eta$.
Evidently, this is independent of the number of propagating modes, $[\eta]$.

%%%%%%%%% FIG. %%%%%%%%%%%%%%
\begin{figure}
%\centerline{\psfig{figure=algaas.eps,width=7cm}}
%\par
\vspace*{5mm}
\caption{\label{fig:gaas}
A quantum point contact is defined at the interface between the undoped
AlGaAs spacer layer
and GaAs by the split gate confinement technique.
Noise is caused by the motion of
impurities in the n-doped AlGaAs layer.
}
\end{figure}

\begin{figure}
%\centerline{\psfig{figure=1.eps,width=8.5cm}}
%\par
%\vspace*{5mm}
\caption{\label{fig:1}
Logarithm of noise, $\log_{10} [S(\omega=100\Gamma_{\min})]$
as a function of $\eta$.
The unit of $S(\omega)$ is
$(2e^2V_{sd}T/h)^2$.
The parameter $\eta$ gives the operating
point, i.e.
where on the conductance-gate voltage staircase
the system is centered.
The integer part of $\eta$ is the number of conducting modes.
We calculated $S_a$ as the numerical Fourier transform of
the valid expression for the noise, Eq. (42). Using the traditional
approach, $S_b$ was computed from
Eq. (44, 47-49).
% \footnote{{\underline Yuri:} Please check.}
The minimum fluctuator potential $a_{\min}$ was defined by
$Na_{\min}/E_{\text F}=5\times10^{-3}$
($E_{\text F}$ being the Fermi energy);
$a_{\max}$ by $a_{\max}/a_{\min}= 100$.
}
\end{figure}

\begin{figure}
%\centerline{\psfig{figure=2.eps,width=8.5cm}}
%\par
%\vspace*{5mm}
\caption{\label{fig:2}
Logarithm of noise, $\log_{10} [S(\omega)]$ (in units of
$(2e^2V_{sd}T/h)^2$)
as a function of the logarithm
of dimensionless frequency,
$ \log_{10} (w)$ ($\omega=w\Gamma_{\min}$).
The minimum fluctuator potential $a_{min}$ was defined by
$Na_{\min}/E_{\text F}=5\times10^{-3}$
($E_{\text F}$ being the Fermi energy);
$a_{\max}$ by $a_{\max}/a_{\min}= 100$.
The calculation was done for two different values of $\eta$,
$\eta=2$ where mode number 2 is just about to become propagating,
and $\eta=2.5$.
The numbers indicate the approximate slope $k$ $(S(\omega) \propto
\omega^k)$ at frequency
$\omega=100\Gamma_{\min}$.
}
\end{figure}

\begin{figure}
%\centerline{\psfig{figure=3.eps,width=8.5cm}}
%\par
%\vspace*{5mm}
\caption{\label{fig:3}
Integrated noise intensity per decade
as a function of dimensionless
center frequency $w_c=\omega_c/\Gamma_{\min}$
%($\omega_c$ real frequency)
{}.
The different plots correspond to the same cases as in Fig.~3,
%~\ref{fig:2} PH: Cannot be used put it in manually.
$\eta=2$,
and $\eta=2.5$.
The integration was
performed over the interval $w\in [w_c-0.5,w_c+0.5]$.
If $S(\omega)\propto 1/\omega^{\alpha}$, where $\alpha=1$,
the bars would be of equal height.
}
\end{figure}

\end{document}